# Unification per the Relational Blockworld

*W.M. Stuckey[1] and Michael Silberstein[2,3]*

## Abstract

We use the Relational Blockworld (RBW) interpretation of quantum mechanics to resolve the foundational problems therein. As predicted by Smolin, the resolution of these problems is not independent of the problem of unification and the nature of time. Specifically, RBW requires a theory fundamental to quantum physics in which one must explicitly construct dynamical/diachronic entities (objects obtained via trans-temporal identification) from 'relations'. We use discrete graph theory to propose heuristically the nature of this underlying theory, which is based on a self-consistency criterion for the mutual construct of dynamical/diachronic entities, space and time whence a spatiotemporally discrete action fundamental to the path integral approach to quantum and classical physics. The proposed unification scheme suggests a novel approach to quantum gravity.

[1] Department of Physics, Elizabethtown College, Elizabethtown, PA 17022, stuckeym@etown.edu
[2] Department of Philosophy, University of Maryland, College Park, MD 20742
[3] Department of Philosophy, Elizabethtown College, Elizabethtown, PA 17022, silbermd@etown.edu

## 1. INTRODUCTION

*1.1 Basis of Unification.* Concerning “the foundational problems of quantum mechanics,” Smolin (2006, 9) writes, “This is probably the most serious problem facing modern science,” and (2006, 10), “The problem of quantum mechanics is unlikely to be solved in isolation; instead, the solution will probably emerge as we make progress on the greater effort to unify physics.” Indeed, the Relational Blockworld (RBW) interpretation of quantum mechanics (Stuckey *et al.*, 2006, 2007 & 2008; Silberstein *et al.*, 2007 & 2008) suggests (if not demands) the promise of unification based on self consistency, not dynamical law. Exactly what we mean by “self consistency” will be mathematically articulated later, but we can suggest conceptually how the “utterly simple idea” (Wheeler, 1985) of self consistency at the basis of our proposed unification scheme stands to resolve, for example, (1) the many problems of ‘multiplicity’ in dynamical approaches to unification, e.g., the uniqueness problem of fundamental constants in the standard model and (2) the uniqueness problem associated with the form of the fundamental dynamical laws themselves, the very special initial and boundary conditions we call the big bang, etc.

Physics is methodologically reductive via unification and its explanans are required to be unique so that explanation is understood to be a mathematical description without alternatives, thereby constituting determination—that is, it is generally held that fundamental physics has only succeeded in explaining *why* (real explanation) rather than *how* (mere description) IFF the fundamental theory uniquely determines, necessitates, or entails all the other higher-level theories and physical facts. Thus, ‘multiplicity’ at the fundamental levels is “a tremendous embarrassment” (Smolin, 2006, 13) on two counts – we’re not getting “more and more in terms of less and less” (Weinberg, 1992) and we don’t have uniqueness in our explanation. Therefore, we agree with most physicists that the standard model *is not ultimately fundamental* and the best accounts of quantum gravity (QG) have failed so far to *explain* the standard model. However, we differ with most physicists in our proposed nature of this fundamental theory. At the foundation of our unification scheme resides a discrete action[1] constructed per a self-consistency criterion (SCC) for dynamical/diachronic entities (objects obtained via trans-temporal identification), space and time. The SCC constrains, among other things, the constants in the standard model *as a set, not in isolation*. Since the hierarchy is "SCC → action → standard model," 'multiplicity' found in the *non*-fundamental standard model is the result of freedom associated with the unique SCC, which *is* fundamental. Our approach constitutes a *unification* of physics as opposed to a mere *discrete approximation* thereto, since we are proposing a source for the action, which is otherwise fundamental.

We believe an SCC of the sort we propose at the foundations of physics might satisfy Wheeler's criterion, i.e., the "utterly simple idea" at the foundations of physics must be compelling (Wheeler, 1985). One reason we find our SCC a compelling basis for physics is as follows. There are only three options for outcomes in the structure of attempted unification in a reductive methodology – there is a foundation at which the project of explanation naturally comes to an end, there is an infinite regress as one seeks the bottom or the structure is (viciously) circular. Virtually all physicists would only count the first option as a victory (cf., Weinberg's "Dreams of a Final Theory," 1993), but here again there is a potential problem. Since reductive explanation requires 'downward' description, the fundamental axioms of a reductive formalism are themselves inexplicable per the formalism proper. As Davies (2007) puts it, "where do the laws of nature come from?" Why do the fundamental laws have the form that they do? Thus, a unified theory constructed per the first option is only ultimately credible if its axiomatic basis is 'self-explanatory' or self-evident. Our proposed unification scheme is a reductive methodology of the first type and, as we will show, the demand for the self-consistency of dynamical/diachronic entities, space and time in our approach is self-evident. Concerning this option, Redhead writes (1990, 152),

> The ideal of scientific explanation is a matter of logical deduction, given a unified set of deep explanatory principles that are themselves accepted, for the time being, without explanation. But of course the ideal of scientific explanation is one for ongoing improvement. Perhaps from the fundamental laws of microphysics, by some consistency criterion, it will turn out that the constants of nature are

[1] Actually, the SCC determines the *invariant core* of the discrete action, i.e., that part of the discrete action which does not vary in the path integral computation of the transition amplitude. We call this invariant core the *actional* or the *kernel of the discrete action*.

> tightly constrained or even uniquely determined. But even then we would still have the task of explaining the laws themselves at a still more fundamental level. At some stage scientific explanations always turn into descriptions—'that's how it is folks'—there is no *ultimate* terminus in science for the awkward child who persists in asking why! I do not believe the aim of some self-vindicating *a priori* foundation for science is a credible one.

We agree with Redhead that explanation per physics is, at the end of the day, description, but he doesn't qualify that claim by saying description *as constrained by* "determination," so the connotation is that of "mere contingent description." Likewise, he suggests a fundamental "consistency criterion" but assumes it would only constrain "constants of nature" and would not serve to explain the form of the laws. Our SCC avoids these complaints because it does not "merely" constrain the constants of nature but provides a self-evident basis for the *action itself*, which Toffoli (2003) notes is necessary to provide a satisfactory foundational explanation since the *action itself* is not "self evident" or "self explanatory."

Exactly how our SCC provides a self-evident basis for our approach to unification will be seen in the construct of a discrete action, but we already have an ideal example in Einstein's equations of general relativity (GR). Momentum, force and energy all depend on spatiotemporal measurements (tacit or explicit), so the stress-energy tensor cannot be constructed without tacit or explicit knowledge of the spacetime metric (technically, the stress-energy tensor can be written as the functional derivative of the matter-energy Lagrangian with respect to the metric). But, if one wants a 'dynamic' spacetime in the parlance of GR, the spacetime metric must depend on the matter-energy distribution in spacetime. GR solves this dilemma by demanding the stress-energy tensor be 'consistent' with the spacetime metric per Einstein's equations. This self-consistency hinges on divergence-free sources, which finds a mathematical counterpart in the topological maxim, "the boundary of a boundary is zero" (Misner *et al.*, 1973). So, Einstein's equations of GR provide an example of an SCC. In fact, our SCC is based on the same topological maxim for the same reason, as are quantum and classical electromagnetism (Misner *et al.*, 1973; Wise, 2006). Thus, we believe the SCC at the bottom of our unification scheme satisfies Wheeler's criterion.

*1.2 Outline of the Paper.* We begin in section 2 with a short explanation of how our new interpretation of QM resolves "the most serious problem facing modern science." According to RBW, relations (not *things*) are the fundamental constituents in a spatiotemporally holistic description of reality, so QM detector clicks are not necessarily evidence of microscopic dynamical/diachronic entities (with "thusness" as Einstein would say) propagating through space and impinging on the detector. Rather, detector clicks can evidence rarefied subsets of relations comprising the source, detector, beam splitters, mirrors, etc. in the experimental arrangement (as in the case of entanglement). In short, RBW provides a fundamentally kinematic (pre-dynamical) account of QM that resolves all the foundational issues therein. Because it is "relations all the way down" in RBW and because our account is foundationally kinematic, RBW is contingent upon a new, spatiotemporally holistic approach to the unification of physics that would provide a fundamentally discrete theory for the construct of dynamical/diachronic entities, space and time via 'relations'.

We explicate our approach to unification in section 3. We believe discrete graph theory provides the best approach to our brand of unification and, since our goal is not to develop new calculational techniques for quantum physics but merely to exploit the formalism conceptually, we keep the examples simple. We begin with a spatially and temporally discrete path integral formalism which yields QM in the spatially discrete, temporally continuous limit of rarefied relations, and quantum field theory (QFT) in the spatially and temporally continuous limits of rarefied relations. The process by which one produces a spatially discrete, temporally continuous action from a spatiotemporally discrete action illustrates the fundamentality of relations and requires explicit trans-temporal identification, which involves the fundamental concepts of time, space and objecthood[2] (see Figure 1). Consequently, we believe the articulation of the otherwise tacit construct of dynamical/diachronic entities has a mathematical counterpart fundamental to the action, which is in accord with Toffoli's belief that there exists a mathematical tautology fundamental to the action (Toffoli, 2003):

> Rather, the motivation is that principles of great generality must be by their very nature *trivial*, that is, expressions of broad tautological identities. If the principle

[2] This is in accord with Smolin's belief that "the nature of time" is "the key" to quantum gravity (Smolin, 2006, 256).

> of least action, which is so general, still looks somewhat mysterious, that means we still do not understand what it is really an expression of—what it is trying to tell us.

We argue that this mathematical counterpart to the process of trans-temporal identification is an SCC uniting the concepts of space, time and dynamical/diachronic entities. In fact, the SCC is our counterpart to "quasiseparability" per Albrecht and Iglesias (2008), i.e., the existence of 'things' suffices to break the symmetry of an otherwise ambiguous solution space and provide a unique set of laws for our universe. Using graph theory *a la* Wise (2006) we note that $\partial_1\partial_1^+$ , where $\partial_1$ is a boundary operator in the chain complex of our simple 2D graph satisfying $\partial_1\partial_2 = 0$, has precisely the same form as the matrix operator in the discrete action for coupled harmonic oscillators. Therefore, we are led to speculate that $\vec{\vec{A}} \propto \partial_1\partial_1^+$ , where $\vec{\vec{A}}$ is the matrix operator of the spatiotemporally discrete action in general. Defining the discrete source vector $\vec{J}$ relationally via links of the graph per $\vec{J} \propto \partial_1\vec{e}$ then gives $\vec{\vec{A}} \cdot \vec{v} \propto \vec{J}$ , where $\vec{e}$ is the vector of links and $\vec{v}$ is the vector of vertices. $\vec{\vec{A}} \cdot \vec{v} \propto \vec{J}$ constitutes what is meant by the "self-consistency" of space, time and dynamical/diachronic objects (the SCC) and thereby supplies $\vec{\vec{A}}$ and $\vec{J}$ for the actional. Thus, we have a basis for the action in accord with Toffoli, since $\vec{\vec{A}} \cdot \vec{v} \propto \vec{J}$ follows from a discrete version of Wheeler's mathematical tautology "the boundary of a boundary is zero," i.e., $\partial_1\partial_2 = 0$. In this approach, Wheeler's "utterly simple idea" is the SCC based on Wheeler's topological maxim (Figure 1).

To further clarify the holistic nature of space, time and 'sources' in this view we compare the relevant (off-diagonal) matrix elements of the spatially discrete, temporally continuous twin-slit amplitude to the Schrödinger twin-slit wavefunction. Per Feynman, the twin-slit experiment "has in it the heart of quantum mechanics. In reality, it contains the *only* mystery" (Feynman *et al.*, 1965). Our comparison illustrates how spatial distance per Schrödinger wave mechanics exists only between interacting sources, as opposed to the common view that spatial distance exists in source-free regions of the spacetime manifold. Thus, in resolving the "mystery" of the twin-slit experiment, RBW satisfies

Pauli's admonition that "in providing a systematic foundation for quantum mechanics, one should start more from the composition and separation of systems than has until now (with Dirac, e.g.) been the case" (Pauli, 1985). And, our rendition of the twin-slit experiment necessarily circumvents "a fundamental incompatibility between general relativity and quantum mechanics" (Howard, 1997, 122), i.e., QM embodies non-separability via quantum entanglement while the metric of general relativity and its underlying differentiable manifold embody pervasive spatiotemporal separability. This is also consistent with the fact that "There are hence no observables of the form of the value of some field at a given point of a manifold, *x*" (Smolin, 2001, 5).

We conclude section 3 by emphasizing that RBW conforms to realism per RWOT (Smolin's "real world out there"). While RBW does not employ human observation or consciousness to solve the measurement problem, it also does not divide "the world into system and observer" (Smolin, 2006, 8), where "observer" means any sort of "detector." Its dependence upon unification notwithstanding, RBW is actually rather simple if one has truly transcended the idea that the dynamical or causal perspective is the most fundamental one. Per RBW, reality is fundamentally described holistically in space and time with relations as the fundamental constituents, and this reality subsumes and exceeds that of dynamism with its temporally forward causal descriptions respecting the common cause principle, i.e., every systematic correlation between events is due to a cause that they share.

As an example of how a spatiotemporally holistic perspective can 'artificially' harbor dynamical 'mystery', one needs to appreciate the blockworld (BW) perspective (Geroch, 1978), i.e.,

> There is no dynamics within space-time itself: nothing ever moves therein; nothing happens; nothing changes. In particular, one does not think of particles as moving through space-time, or as following along their world-lines. Rather, particles are just in space-time, once and for all, and the world-line represents, all at once, the complete life history of the particle.

When Geroch says that "there is no dynamics within space-time itself," he is not denying that the mosaic of the blockworld possesses patterns that can be described with dynamical laws. Nor is he denying the predictive and explanatory value of such laws. Rather, given the reality of all events in a blockworld, dynamics are not "event factories" that bring heretofore non-existent events (such as measurement outcomes) into being. Dynamical laws are not brute unexplained explainers that "produce" events. Geroch is advocating for what philosophers call Humeanism about laws. Namely, the claim is that dynamical laws are *descriptions of regularities* and not the *brute explanation* for such regularities. His point is that in a blockworld, Humeanism about laws is an obvious position to take because everything is just "there" from a "God's eye" (Archimedean) point of view. That is, all events past, present and future are equally "real" in a blockworld. Thus, in the blockworld view, the existence of any particular point of the spacetime manifold is no more or less mysterious than that of any other. Whereas, if we insist that the dynamical view is fundamental, then we demand that each spatial hypersurface be explained by conditions on the spatial hypersurface immediately preceding it, leaving an initial hypersurface 'unexplained'. In this manner, we've created a 'faux' (from the BW perspective) "mystery," e.g., what *caused* the big bang? Likewise, attempting to explain all QM phenomena via dynamism precludes certain blockworld descriptions rendered by RBW (e.g., Stuckey *et al.*, 2008). Thus, the dynamical perspective is overly constrained because it constitutes a proper subset of all possible BW explanations. "Mysterious" QM phenomena are totally explicable via RWOT, but dynamical reality is only a proper subset of a spatiotemporally holistic reality and some QM phenomena are "mysterious" simply because they're not elements of that dynamical subset.

In order to avoid trivializing the BW explanation, BW interpretations of QM invoke clever devices such as time-like backwards causation (Price, 1996), advanced action (Cramer, 1986) and the two-vector formalism (Aharonov *et al.*, 1964; Elitzur & Vaidman, 1993). Do these beautiful and clever devices really avoid the charge of triviality? Such explanations are no less *dynamical* than standard quantum mechanics, which is puzzling given that the original BW motivation for such accounts lacks *absolute* change and becoming. As far we know, only Cramer speaks to this worry. Cramer notes (1986, 661) that the backwards-causal elements of his theory are "only a pedagogical convention," and that in fact "the process is atemporal." Indeed, it seems to us that all such dynamical or causal devices in a BW should be viewed fundamentally as book keeping. Backwards causation quantum mechanics (BCQM) and the like, even having acknowledged the potential explanatory importance of BW, have not gone far enough in their atemporal, acausal and adynamical thinking. Whereas such accounts are willing to think backwardly, temporally speaking, it is still essentially *dynamical, temporal* thinking.

We rather believe the key to rendering a BW explanation nontrivial is to provide an algorithm for the relevant BW construction. Thus, the answer to “Why did X follow Y and Z?” is not merely, “Because X is already ‘there’ in the future of Y and Z per the blockworld,” but as we will illustrate, “Because this must be the spatiotemporal relationship of X, Y and Z in the blockworld per the self-consistent definition of the entities involved in X, Y and Z.” If one chooses to read dynamical stories from a BW picture, they may where feasible. However, BW descriptions are not limited to the depiction of dynamical/causal phenomena, so they are not constrained to dynamical/causal storytelling. In the following passage, Dainton (2001) paints a suggestive picture of what it means to take the BW perspective seriously both ontologically and explanatorily:

> Imagine that I am a God-like being who has decided to design and then create a logically consistent universe with laws of nature similar to those that obtain in our universe…Since the universe will be of the block-variety I will have to create it *as a whole*: the beginning, middle and end will come into being together…Well, assume that our universe is a static block, even if it never ‘came into being’, it nonetheless exists (timelessly) as a coherent whole, containing a globally consistent spread of events. At the weakest level, “consistency” here simply means that the laws of logic are obeyed, but in the case of universes like our own, where there are universe-wide laws of nature, the consistency constraint is stronger: everything that happens is in accord with the laws of nature. In saying that the consistency is “global” I mean that the different parts of the universe all have to fit smoothly together, rather like the pieces of a well-made mosaic or jigsaw puzzle.

Does reality contain phenomena which *strongly suggest* an acausal BW algorithm? According to RBW, the deepest explanation of EPR-Bell correlations is such an algorithm. QM *a la* RBW provides an acausal BW algorithm in its prediction of Bell inequality violations and these violations have been observed. So it appears that reality does harbor acausal BW phenomena and QM *a la* RBW is one algorithm for depicting the self-consistent placement of such phenomena in a blockworld.

The blockworld of RBW is precisely in keeping with Geroch's "all at once" notion of spacetime and Dainton's "vast spatiotemporal mosaic," but it is important to note again that it is a non-separable BW while that of relativity theory is separable. That is to say, the metric field of relativity theory takes on values at each point of the differentiable spacetime manifold, even in regions where the stress-energy tensor is zero, as if "things" are distinct from the concepts of space and time. Per RBW, the concepts of space, time and dynamical/diachronic entities can only be defined self-consistently so each is meaningless in the absence of the others. Accordingly, there need not be an 'exchange' particle or wave moving 'through space' between the worldlines of trans-temporal objects to dynamically mediate their interaction and establish their spatial separation, e.g., clicks 'caused' by "screened off" particles. As a consequence, we understand that a QM detection event (subset of the detector) results from a particular, rarefied subset of the relations defining sources, detectors, beam splitters, mirrors, etc. in an "all at once" fashion. In this picture, there are no "screened off" particles moving in a wave-like fashion through separable elements of the experimental arrangement to cause detection events, but rather detection events are evidence that the experimental equipment itself is non-separable[3]. While non-separable, RBW upholds locality in the sense that there is no action at a distance, no instantaneous dynamical or causal connection between space-like separated events. And, there are no space-like worldlines in RBW. Thus, we have *the non-separability of dynamical entities, e.g., sources and detectors, while the entities themselves respect locality*. In this sense, we agree with Howard (1997, 124) that QM is best understood as violating "separability" (i.e., independence) rather than "locality."

One might perceive a certain tension in the combination of relationalism and the BW perspective. After all, nothing seems more *absolute* than the BW viewed as a whole, hence the Archimedean metaphor. One can just imagine Newton's God gazing upon the timeless, static 4-dimensional BW mosaic (her sensorium) from her perch in the fifth (or higher) dimension; what could be more absolute? But relationalism is a rejection of the absolute and the very idea of a God's eye perspective. In any case, one must never forget that while RBW is a blockworld in the sense that all events are equally real, it is a *relational* blockworld so there is no meaning to a God's eye perspective, i.e., any beings observing the BW must be *a part of it*. Short of occupying all the perspectives "at once," there is nothing that corresponds to such a privileged view.

---

[3] Since space, time and dynamical/diachronic entities are to be mutually and self-consistently defined (via relations), the non-separability of spacetime entails the non-separability of dynamical/diachronic entities and vice-versa. RBW does away with any matter/geometry dualism.

**Figure 1**

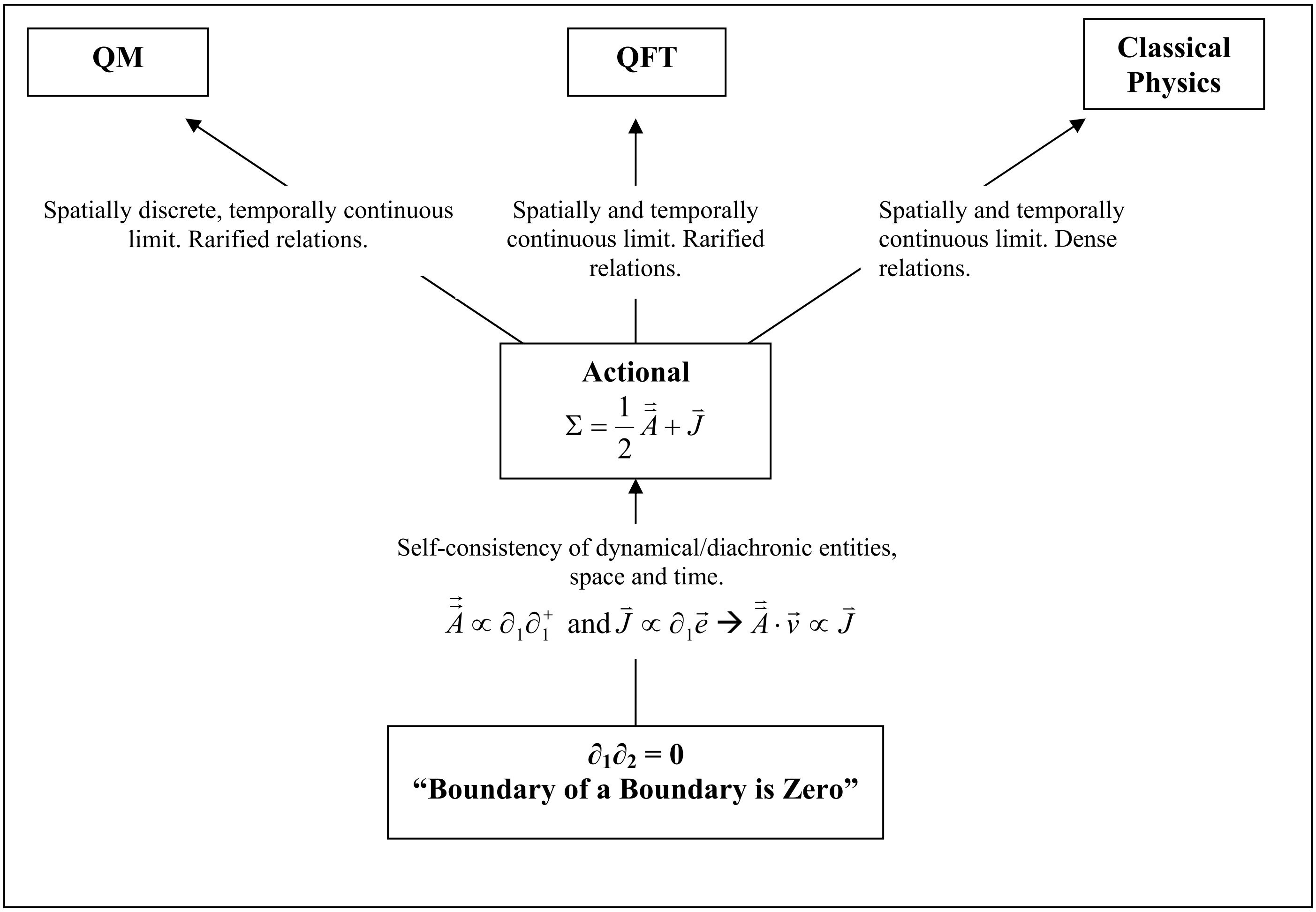

Clearly, therefore, RBW resonates strongly with Smolin's belief that what "we are all missing" in the search for unification "involves two things: the foundations of quantum mechanics and the nature of time" (Smolin, 2006, 256). To summarize RBW's approach to unification, the fundamental principle of physics (SCC) dictates that the actional be constructed from a spatiotemporally holistic graph of divergence-free sources and non-separable fields per the topological maxim "the boundary of a boundary is zero." The description of a particular phenomenon in a particular regime of physics then obtains from the appropriate limit of its action. Regimes currently being explored are the spatially discrete and temporally continuous distribution of rarefied relations (quantum mechanics), the spatially and temporally continuous distribution of rarefied relations (quantum field theory) and the spatially and temporally continuous distribution of dense relations (classical physics). Per this unification scheme, QG would require a mix of classical and quantum regimes, e.g., using spectra (spatially discrete and temporally continuous distribution of rarefied relations) to ascertain stellar kinematics (spatially and temporally continuous distribution of dense relations). We discuss our brand of QG in section 4 and conclude in section 5 with an overview of what we believe merits further scrutiny.

## 2. RESOLVING THE "MOST SERIOUS PROBLEM"

The RBW interpretation of non-relativistic quantum mechanics is thoroughly expounded elsewhere (Stuckey *et al.*, 2006, 2007 & 2008; Silberstein *et al.*, 2007 & 2008), so we provide only the briefest overview here. Per RBW, the spacetime of QM is a relational, non-separable blockworld whereby spatial distance is only defined between interacting trans-temporal objects. RBW is motivated, in part, by a result due to Kaiser (1981), Bohr & Ulfbeck (1995) and Anandan (2003) who showed independently that the non-commutivity of the position and momentum operators in QM follows from the non-commutivity of the Lorentz boosts and spatial translations in special relativity, i.e., the relativity of simultaneity. That relations are fundamental to dynamical/diachronic entities, as opposed to the converse per a dynamic perspective, is motivated by the work of Bohr, Mottelson & Ulfbeck (2004) who showed how the quantum density operator can be obtained via the symmetry group of the relevant observable.

*2.1 The Measurement Problem.* RBW deflates the measurement problem with a novel form of a "statistical interpretation." The fundamental difference between our version of this view and the usual understanding of it is that on the usual view the state description refers to an "ensemble" which is an ideal collection of similarly prepared quantum particles, whereas "ensemble" according to our view is just an ideal collection of spacetime regions $S_i$ "prepared" with the same spatiotemporal boundary conditions per the experimental configuration itself. The union of the click events in each $S_i$, as $i \rightarrow \infty$, produces the characteristic Born distribution. Accordingly, probability in RBW is interpreted per relative frequencies.

The wavefunction description of a quantum system can be interpreted statistically because we understand that, as far as measurement outcomes are concerned, the Born distribution has a basis in the symmetry of the actional for the experimental configuration. Each "click," which some would say corresponds to the impingement of a particle onto a measurement device with probability computed from the wavefunction, corresponds to rarefied relations in the context of the experimental configuration. The measurement problem *exploits* the possibility of extending the wavefunction description from the quantum system to the whole measurement apparatus, whereas the spatiotemporally holistic description of RBW *already includes* the apparatus. The measurement problem is therefore a non-starter on our view. The measurement problem arises because of the assumption that *dynamics* are the deepest part of the explanatory story, an assumption RBW rejects, therefore RBW is a kinematic interpretation of QM.

Since a dynamical/diachronic entity (such as a detector) possesses properties (to include click distributions) according to a spatiotemporally global set of relations (all dynamical/diachronic entities are defined non-separably in "a vast spatiotemporal mosaic"), one could think of RBW as a local hidden-variable theory (such as BCQM) whereby the relations or symmetries provide the "hidden variables." One can construct a *local* hidden-variable theory if one is willing to claim that systems which presumably have not interacted may nevertheless be correlated. Such correlations appear to require some kind of universal conspiracy behind the observed phenomena, hence Peter Lewis (2008) calls such theories "*conspiracy theories.*" As he says, "the obvious strategy is the one that gives conspiracy theories their name; it involves postulating a vast, hidden mechanism whereby systems that apparently have no common past may nevertheless have interacted." Independence is the assumption that the hidden variables assigned to the particles are independent of the settings of the measuring devices. If Independence is violated, then a local hidden-variable theory (a conspiracy theory) can in principle account for the Bell correlations. But how *could* Independence be violated? The common cause principle tells us that every systematic correlation between events is due to a cause that they share. As a trivial consequence, systems that have not interacted cannot be systematically correlated, and all appearances indicate that the particles and the measuring devices in EPR-Bell phenomena do not interact before the measurement. Lewis (2008) suggests three possibilities for violating Independence:

> Hidden-mechanism theories and backwards-causal theories are both strategies for constructing a local hidden-variable theory by violating Independence. The first of these postulates a mechanism that provides a cause in the past to explain the Bell correlations, and the second postulates a cause in the future. But there is a third strategy that is worth exploring here, namely that the common cause principle is *false*—that some correlations simply require no causal explanation.

Lewis calls the third strategy of denying the common cause principle "acausal conspiracy theories;" RBW can be reasonably characterized in this fashion with the symmetry of the actional playing the role of the hidden-variable. However such a characterization is also misleading in that we are not supplementing QM in any standard sense, such as modal interpretations *a la* Bohm. We are not claiming that quantum mechanics is incomplete but that the symmetry of the actional provides a deeper explanation than QM as standardly and *dynamically* conceived. At least at this level, there is no deeper explanation for individual outcomes of quantum experiments than that provided by the symmetry of the actional underlying each experimental configuration.

*2.2 Entanglement and Non-locality.* The spatiotemporally holistic description of the experimental configuration includes the experimental outcomes, and it is possible that those outcomes are correlated via the SCC. Since the actional—constructed to represent a specific subset of reality instantiated (approximately) by the experiment in question—is "all at once" to include outcomes, there is no reason to expect entanglement will respect any kind of common cause principle. As we stated *supra*, causality/dynamism are not essential in the algorithm for constructing a blockworld description. Although RBW is *fundamentally* non-dynamical (relata from relations "all at once," rather than relata from relata in a causal or dynamical structure), it does not harbor non-locality in the odious sense of "spooky action at a distance" as in Bohm for example, i.e., there are no space-like worldlines (implied or otherwise) between space-like separated, correlated outcomes. Again, this is where RBW suggests a new approach to fundamental physics because dynamical/diachronic entities are modeled fundamentally via relations in "a vast spatiotemporal mosaic" instead of via "interacting" dynamical constituents *a la* particle physics[4].

Our account provides a clear description, in terms of relations in a blockworld, of quantum phenomena *that does not suggest the need for a "deeper" causal or dynamical explanation*. If explanation is simply determination, then our view explains the structure of quantum correlations by invoking what can be called *acausal, adynamical global determination relations*. In QM, these "all at once" determination relations are given by the actional which underlies a particular experimental set-up. Not objects governed by dynamical laws, but rather acausal relations per the relevant actional do the fundamental explanatory work according to RBW. We can invoke the *entire* spacetime configuration of the experiment so as to predict, and explain, the EPR-Bell correlations. Therefore, RBW provides a geometrical, acausal and adynamical account of entanglement.

**Table 1**

| **Changing the *Fundamental* Explanatory Model** | |
|---|---|
| Dynamical model of explanation: | RBW model of explanation: |
| Fundamental constituents harbor (tacit) trans-temporal identity (particles, fields, waves, etc.). | Fundamental constituents are relations. |
| Continuous action constructed per relevant symmetries gives | Actional obtained per self-consistency of dynamical/diachronic entities, space and time. |
| Dynamical equations of motion. These plus | Amplitude for specific spatiotemporal configuration computed via appropriate limit of discrete action in "path integral." |
| Initial/boundary conditions | Probability = amplitude squared. |
| Determine evolution in 4-space or configuration space (which determines evolution in 4-space). | Find normalized probabilities for all possibilities. |
| = fundamental explanation | = fundamental explanation |

4 This means particles physics per QFT is displaced from its fundamental status (Figure 1).

## 3. THE PROPOSED UNIFICATION SCHEME

*3.1 Discrete Path Integral Formalism.* If RBW is to be accepted as a viable interpretation of QM, it is incumbent upon us to provide, at very least, a heuristic approach to unification whereby space, time and diachronic objects are mutually and relationally defined. In response to this challenge, we use graph theory since it has already been shown to provide an excellent mathematics for the construct of a discrete basis to quantum physics (e.g., Markopoulou & Smolin, 2004). The manner by which this story underwrites physics as currently practiced is conceptually simple – we obtain three limiting cases roughly corresponding to each of QM, QFT and classical physics. The first limiting case is $\Delta t \rightarrow 0$ to form rarefied collections of worldlines for "sources[5]" while keeping these sources located discretely in space (QM). The second limiting case is $\Delta t \rightarrow 0$ and $\Delta\vec{x} \rightarrow 0$ while maintaining a rarefied collection of sources (QFT). The third limiting case is $\Delta t \rightarrow 0$ and $\Delta\vec{x} \rightarrow 0$ while allowing for a dense distribution of sources (classical physics). This scenario is depicted in Figure 1. In a nutshell, that's how the proposed formalism for RBW makes correspondence with "higher level" physics.

The value of this approach is conceptual, if not analytical, in that it provides a basis for visualizing a relational, "all at once" model of reality fundamental to that of dynamism. We use our graphical approach to generate the kernel for a discrete transition amplitude, $Z$, for sources without scattering, i.e.,

$$Z = \int \ldots \int dQ_1 \ldots dQ_N \exp\left[\frac{i}{2}\vec{Q}\cdot\vec{\vec{A}}\cdot\vec{Q} + i\vec{J}\cdot\vec{Q}\right] \tag{1}$$

(Zee, 2003) whence

$$Z = \left(\frac{(2i\pi)^N}{\det(A)}\right)^{1/2} \exp\left[-\frac{i}{2}\vec{J}\cdot\vec{\vec{A}}^{-1}\cdot\vec{J}\right] \tag{2}.$$

The matrix $\vec{\vec{A}}$ is the spatiotemporally discrete form of the differential operator in the action while $\vec{J}$ is the spatiotemporally discrete form of the 'sources'. $Z$ can be viewed as a measure of the "symmetry" contained its kernel or *actional*

$$\Sigma = \frac{1}{2}\vec{\vec{A}} + \vec{J} \tag{3}$$

[5] In the parlance of quantum field theory, particles are created at one "source," then move through space where they are annihilated at another "source," a.k.a. "sink."

which yields the discrete action after operating on a particular vector $\vec{Q}$, e.g., plugging a different path $x$(t) into the Lagrangian produces a different action. That is, the action is the *range* of the actional, which represents a *fundamental, invariant description of the experimental arrangement in the computation of* Z. For this reason, we call $Z$ the *symmetry amplitude* of the spatiotemporal experimental configuration (to include the outcomes), where "measure of symmetry" is understood broadly as the coherence of stationary points in the phase (action) of $Z$. [Classical physics emerges from the computation of $Z$ *a la* lattice gauge theory → quantum field theory → classical physics.]

Of interest here are the ontological implications of the QM limit, i.e., the QM probability amplitude is obtained in the spatially discrete, temporally continuous limit of $Z$, e.g., $Q_n \rightarrow q_i(t)$. The resulting spatially discrete distribution of interacting sources $J_i(t)$ illustrates a key aspect of RBW ontology, i.e., interaction without mediation – there is an interaction of sources without mediating waves or particles traveling through the intervening space. The spatiotemporally discrete formalism also illustrates nicely how QM tacitly assumes an *a priori* process of trans-temporal identification, $J_n \rightarrow J_i(t)$. Indeed, there is no principle which dictates the construct of diachronic entities fundamental to the formalism of dynamics in general—these objects are "put in by hand." When Albrecht and Iglesias (2008) allowed time to be an "internal variable" after quantization, as in the Wheeler-DeWitt equation, they found "there is no one set of laws, but a whole library of different cosmic law books" (Siegfried, 2008). They called this the "clock ambiguity." In order to circumvent this "arbitrariness in the predictions of the theory" they proposed that "the principle behind the regularities that govern the interaction of entities is … the idea that individual entities exist at all" (Siegfried, 2008). Albrecht and Iglesias (2008) characterize this as "the central role of quasiseparability." Similarly, the RBW approach to unification requires a fundamental principle whence the trans-temporal identity employed tacitly in QM and all dynamical theories. Our graphical starting point does not contain dynamical/diachronic entities, space or time, *per se* so we must formalize counterparts to these concepts. Clearly, the process $J_n \rightarrow J_i(t)$ is an organization of the set $J_n$ on two levels—there is the split of the set into $i$ subsets, one for each 'source', and there is the ordering $t$ over each subset. The split represents space (true multiplicity from apparent identity), the ordering represents time (apparent identity from true multiplicity)[6] and the result is objecthood. In this sense, space, time and 'things' are inextricably linked in our formalism, so we suggest they be related self-consistently. We next propose a method for articulating this self-consistency criterion (SCC) at the basis of physics. As it turns out, the principle behind our SCC is none other than Wheeler's "boundary of a boundary is zero" (Misner *et al.*, 1973), which is already employed widely in physics to ensure the consistency of divergence-free sources with their relevant fields.

We will consider a simple graph with six vertices, seven links and two plaquettes (cells) for our 2D spacetime model (Figure 2). Our goal is not to develop new calculational techniques for quantum physics (although graph theory can be used for that purpose), but to exploit the formalism conceptually, i.e., to illustrate what it might mean to mutually define space, time and diachronic objects via a self-consistency criterion. We begin by constructing the boundary operators over our graph.

**Figure 2**

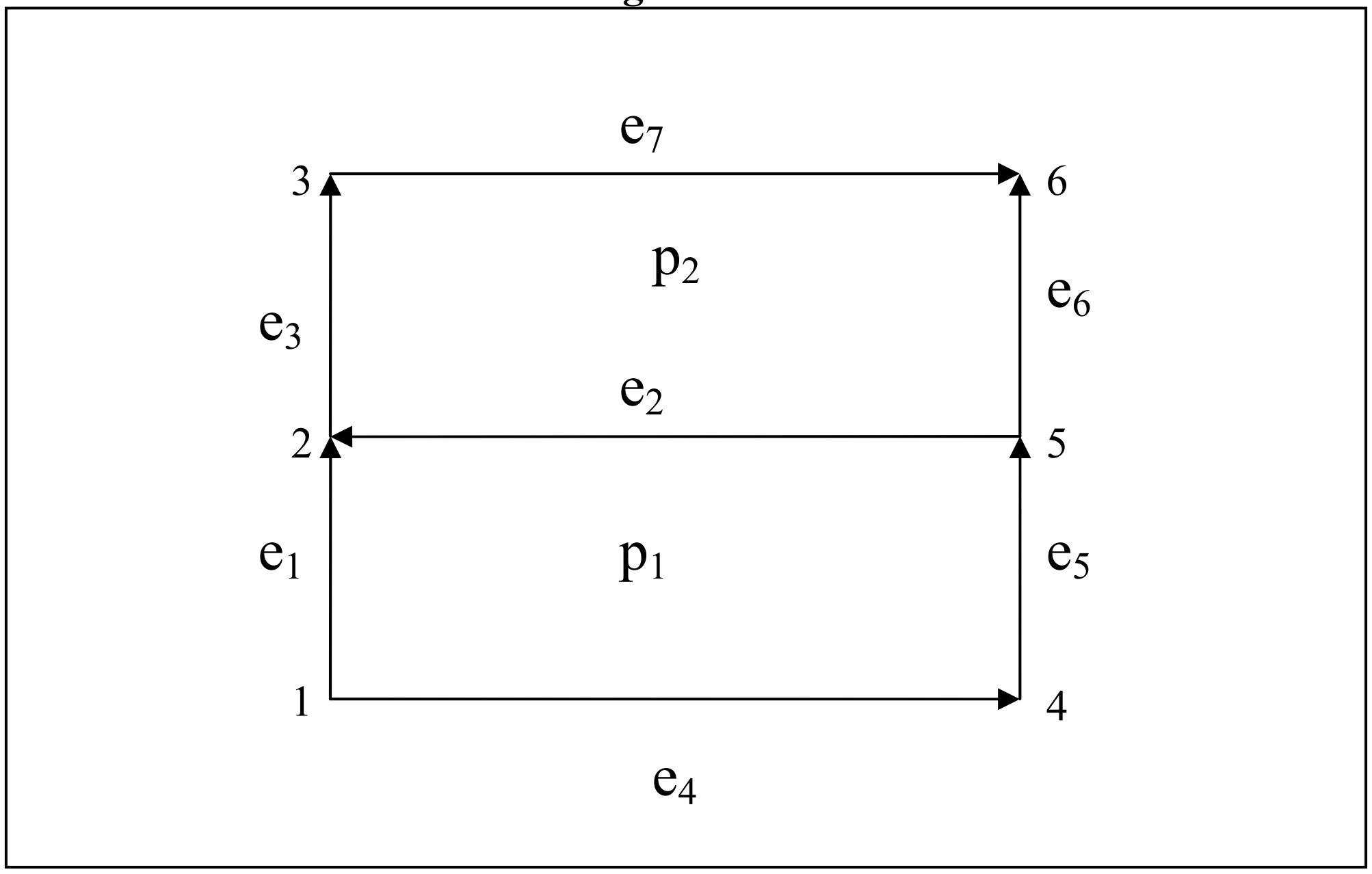

The boundary of $\mathbf{p}_1$ is $\mathbf{e}_4 + \mathbf{e}_5 + \mathbf{e}_2 - \mathbf{e}_1$, which also provides an orientation. The boundary of $\mathbf{e}_1$ is $\mathbf{v}_2 - \mathbf{v}_1$, which likewise provides an orientation. Using these conventions for the orientations of links and plaquettes we have the following boundary operator for

[6] These definitions of space and time follow from a fundamental principle of standard set theory, *multiplicity iff discernibility* (Stuckey, 1999).

$C_2 \rightarrow C_1$, i.e., space of plaquettes mapped to space of links in the spacetime chain complex:

$$\partial_2 = \begin{bmatrix} -1 & 0 \\ 1 & -1 \\ 0 & -1 \\ 1 & 0 \\ 1 & 0 \\ 0 & 1 \\ 0 & -1 \end{bmatrix} \tag{4}$$

Notice the first column is simply the links for the boundary of $\mathbf{p}_1$ and the second column is simply the links for the boundary of $\mathbf{p}_2$. We have the following boundary operator for $C_1 \rightarrow C_0$, i.e., space of links mapped to space of vertices in the spacetime chain complex:

$$\partial_1 = \begin{bmatrix} -1 & 0 & 0 & -1 & 0 & 0 & 0 \\ 1 & 1 & -1 & 0 & 0 & 0 & 0 \\ 0 & 0 & 1 & 0 & 0 & 0 & -1 \\ 0 & 0 & 0 & 1 & -1 & 0 & 0 \\ 0 & -1 & 0 & 0 & 1 & -1 & 0 \\ 0 & 0 & 0 & 0 & 0 & 1 & 1 \end{bmatrix} \tag{5}$$

which completes the spacetime chain complex, $C_0 \xleftarrow{\partial_1} C_1 \xleftarrow{\partial_2} C_2$. Notice the columns are simply the vertices for the boundaries of the edges. These boundary operators satisfy $\partial_1\partial_2 = 0$ as required for "boundary of a boundary is zero," in accord with divergence-free sources (Misner *et al.*, 1973). We want our SCC ultimately founded on this topological maxim so we construct our actional from the boundary operators of our spacetime chain complex. The manner by which we do this is suggested by the discrete action for coupled harmonic oscillators on our simple graph.

The potential for coupled oscillators can be written

$$V(q_1, q_2) = \sum_{a,b} \frac{1}{2} k_{ab} q_a q_b = \frac{1}{2} k q_1^2 + \frac{1}{2} k q_2^2 + k_{12} q_1 q_2 \tag{6}$$

where $k_{11} = k_{22} = k$ (positive) and $k_{12} = k_{21}$ (negative) per the classical analogue (Figure 3)

**Figure 3**

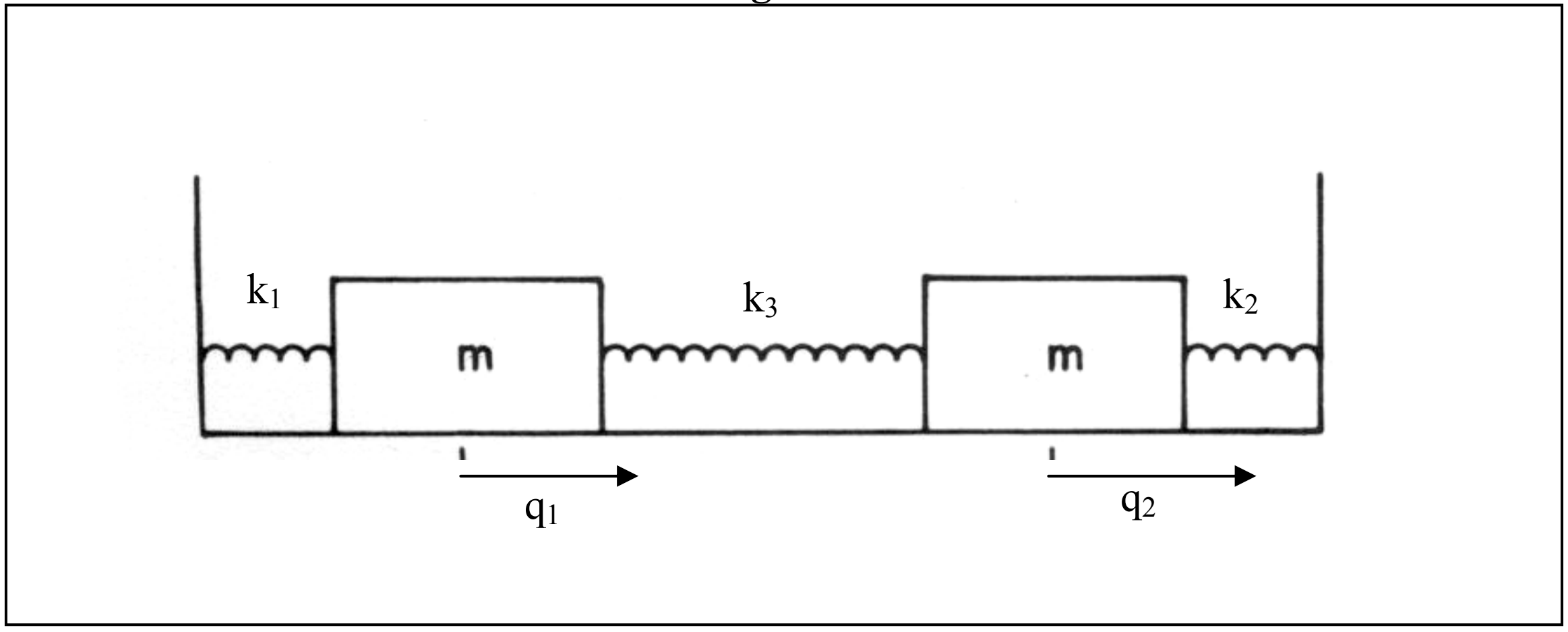

with k = $k_1 + k_3 = k_2 + k_3$ and $k_{12} = -k_3$ to recover the form in Eq. (6). The Lagrangian is then

$$L = \frac{1}{2}m\dot{q}_1^2 + \frac{1}{2}m\dot{q}_2^2 - \frac{1}{2}kq_1^2 - \frac{1}{2}kq_2^2 - k_{12}q_1q_2 \tag{7}$$

so our QM symmetry amplitude is

$$Z = \int Dq(t)\exp\left[-\int_0^T dt\left[\frac{1}{2}m\dot{q}_1^2 + \frac{1}{2}m\dot{q}_2^2 + V(q_1,q_2) - J_1q_1 - J_2q_2\right]\right] \tag{8}$$

after Wick rotation. This gives

$$\vec{\vec{A}} = \begin{bmatrix} \left(\frac{m}{\Delta t} + k\Delta t\right) & \frac{-m}{\Delta t} & 0 & k_{12}\Delta t & 0 & 0 \\ \frac{-m}{\Delta t} & \left(\frac{2m}{\Delta t} + k\Delta t\right) & \frac{-m}{\Delta t} & 0 & k_{12}\Delta t & 0 \\ 0 & \frac{-m}{\Delta t} & \left(\frac{m}{\Delta t} + k\Delta t\right) & 0 & 0 & k_{12}\Delta t \\ k_{12}\Delta t & 0 & 0 & \left(\frac{m}{\Delta t} + k\Delta t\right) & \frac{-m}{\Delta t} & 0 \\ 0 & k_{12}\Delta t & 0 & \frac{-m}{\Delta t} & \left(\frac{2m}{\Delta t} + k\Delta t\right) & \frac{-m}{\Delta t} \\ 0 & 0 & k_{12}\Delta t & 0 & \frac{-m}{\Delta t} & \left(\frac{m}{\Delta t} + k\Delta t\right) \end{bmatrix} \tag{9}$$

on our graph. Thus, we borrow (loosely) from Wise (2006) and suggest $\vec{\vec{A}} \propto \partial_1\partial_1^+$ since

$$\partial_1 \partial_1^+ = \begin{bmatrix} 2 & -1 & 0 & -1 & 0 & 0 \\ -1 & 3 & -1 & 0 & -1 & 0 \\ 0 & -1 & 2 & 0 & 0 & -1 \\ -1 & 0 & 0 & 2 & -1 & 0 \\ 0 & -1 & 0 & -1 & 3 & -1 \\ 0 & 0 & -1 & 0 & -1 & 2 \end{bmatrix} \quad (10)$$

produces precisely the same form as Eq. (9) and quantum theory is known to be "rooted in this harmonic paradigm" (Zee, 2003, 5). [In fact, these matrices will continue to have the same form as one increases the number of vertices in Figure 2.] Now we construct a suitable candidate for $\vec{J}$, relate it to $\vec{\vec{A}}$ and infer our SCC.

Recall that $\vec{J}$ has a component associated with each node so here it has components, $J_n$, n = 1, 2, …, 6; $J_n$ for n = 1, 2, 3 represents one 'source' and $J_n$ for n = 4, 5, 6 represents another 'source'. We propose $\vec{J} \propto \partial_1 \vec{e}$ ,where $e_i$ are the links of our graph, since

$$\partial_1 \vec{e} = \begin{bmatrix} -1 & 0 & 0 & -1 & 0 & 0 & 0 \\ 1 & 1 & -1 & 0 & 0 & 0 & 0 \\ 0 & 0 & 1 & 0 & 0 & 0 & -1 \\ 0 & 0 & 0 & 1 & -1 & 0 & 0 \\ 0 & -1 & 0 & 0 & 1 & -1 & 0 \\ 0 & 0 & 0 & 0 & 0 & 1 & 1 \end{bmatrix} \begin{bmatrix} e_1 \\ e_2 \\ e_3 \\ e_4 \\ e_5 \\ e_6 \\ e_7 \end{bmatrix} = \begin{bmatrix} -e_1 - e_4 \\ e_1 + e_2 - e_3 \\ e_3 - e_7 \\ e_4 - e_5 \\ -e_2 + e_5 - e_6 \\ e_6 + e_7 \end{bmatrix} \quad (11)$$

provides a means of understanding vertices in terms of links and ultimately we want sources defined relationally. For example, vertex 1 is the origin of both links 1 and 4 (negative/positive means the link starts/ends at that vertex) and the first entry of $\partial_1 \vec{e}$ is $-e_1 - e_4$. Since $J_n$ are associated with the vertices to represent 'things', $\vec{J} \propto \partial_1 \vec{e}$ is a graphical representation of "relata from relations." [Note: $\partial_1 \vec{e}$, which we denote $\vec{v}\,*$ and associate with $\vec{v}$, is not equal to $\vec{v}$ proper as will be seen below.]

With these definitions of $\vec{\vec{A}}$ and $\vec{J}$ we have, *ipso facto*, $\vec{\vec{A}}\vec{v} \propto \vec{J}$ as the basis of our SCC since

$$\partial_1\partial_1^+\vec{v} = \begin{bmatrix} 2 & -1 & 0 & -1 & 0 & 0 \\ -1 & 3 & -1 & 0 & -1 & 0 \\ 0 & -1 & 2 & 0 & 0 & -1 \\ -1 & 0 & 0 & 2 & -1 & 0 \\ 0 & -1 & 0 & -1 & 3 & -1 \\ 0 & 0 & -1 & 0 & -1 & 2 \end{bmatrix}\begin{bmatrix} v_1 \\ v_2 \\ v_3 \\ v_4 \\ v_5 \\ v_6 \end{bmatrix} = \begin{bmatrix} 2v_1 - v_2 - v_4 \\ -v_1 + 3v_2 - v_3 - v_5 \\ -v_2 + 2v_3 - v_6 \\ -v_1 + 2v_4 - v_5 \\ -v_2 - v_4 + 3v_5 - v_6 \\ -v_3 - v_5 + 2v_6 \end{bmatrix} = \begin{bmatrix} -e_1 - e_4 \\ e_1 + e_2 - e_3 \\ e_3 - e_7 \\ e_4 - e_5 \\ -e_2 + e_5 - e_6 \\ e_6 + e_7 \end{bmatrix} = \partial_1\vec{e} = \vec{v}^* \qquad (12)$$

where we've used $e_1 = v_2 - v_1$ (etc.) to obtain the last column, which constitutes a definition of links in terms of vertices. It is clear that according to this definition of links in terms of vertices, $\vec{v} \neq \vec{v}^*$ ($\partial_1\partial_1^+$ is not the identity matrix). In fact we have

$$\partial_1^+\vec{v} = \begin{bmatrix} -1 & 1 & 0 & 0 & 0 & 0 \\ 0 & 1 & 0 & 0 & -1 & 0 \\ 0 & -1 & 1 & 0 & 0 & 0 \\ -1 & 0 & 0 & 1 & 0 & 0 \\ 0 & 0 & 0 & -1 & 1 & 0 \\ 0 & 0 & 0 & 0 & -1 & 1 \\ 0 & 0 & -1 & 0 & 0 & 1 \end{bmatrix}\begin{bmatrix} v_1 \\ v_2 \\ v_3 \\ v_4 \\ v_5 \\ v_6 \end{bmatrix} = \begin{bmatrix} v_2 - v_1 \\ v_2 - v_5 \\ v_3 - v_2 \\ v_4 - v_1 \\ v_5 - v_4 \\ v_6 - v_5 \\ v_6 - v_3 \end{bmatrix} = \begin{bmatrix} e_1 \\ e_2 \\ e_3 \\ e_4 \\ e_5 \\ e_6 \\ e_7 \end{bmatrix} = \vec{e} \qquad (13)$$

which reflects the statement *supra*, "Notice the columns are simply the vertices for the boundaries of the edges." Thus, the SCC $\vec{\vec{A}}\vec{v} \propto \vec{J}$ obtains tautologically via the maxim, "the boundary of a boundary is zero," as desired.

Using $\vec{J} = \alpha\partial_1\vec{e}$ and $\vec{\vec{A}} = \beta\partial_1\partial_1^+$ with the SCC gives $\vec{\vec{A}}\vec{v} = \frac{\beta}{\alpha}\vec{J}$, so that $\vec{v} = \frac{\beta}{\alpha}\vec{\vec{A}}^{-1}\vec{J}$. However, $\vec{\vec{A}}^{-1}$ doesn't exist because $\vec{\vec{A}}$ is singular, which means of course that Eq. (1) is ill-defined for this problem. The reason $\vec{\vec{A}}$ is singular is because one of its eigenvalues is zero and that obtains because $\vec{\vec{A}}$ is a difference matrix whose rows are vectors spanning an (N-1)-dimensional hyperplane of the N-dimensional vector space. [This is germane to dynamical difference matrices.] The eigenvector with eigenvalue of zero is normal to this hyperplane (which you can see is [1,1,1,…,1] since $\sum_j A_{ij} = 0$), so we propose a discrete 'renormalization' of Eq (1) whereby the integral is restricted to the hyperplane containing the vectors of $\vec{\vec{A}}$, i.e.,

$$Z = \int_{-\infty}^{\infty} \dots \int_{-\infty}^{\infty} d\tilde{Q}_1 \dots d\tilde{Q}_{N-1} \exp\left[\sum_{j=1}^{N-1}\left(\frac{i}{2}\tilde{Q}_j^2 a_j + i\tilde{J}_j\tilde{Q}_j\right)\right] \tag{14}$$

where $\tilde{Q}_j$ are the coordinates associated with the eigenbasis of $\vec{\vec{A}}$ and $\tilde{Q}_N$ is associated with eigenvalue zero, $a_j$ are the eigenvalues of $\vec{\vec{A}}$ corresponding to $\tilde{Q}_j$, and $\tilde{J}_j$ are the components of $\vec{J}$ in the eigenbasis of $\vec{\vec{A}}$. Again, on our view, $Z$ does not reflect a "sum over all paths in configuration space," but rather it is a 'mathematical machine' which produces a relative symmetry measure of the various Σ associated with different experimental outcomes and configurations. Since $\vec{J}$ resides in this (N-1)-dimensional hyperplane as well (which you can see from $\sum_i J_i = 0$), restricting the integral in Eq. (1) to the hyperplane spanned by the vectors of $\vec{\vec{A}}$ is not at all unreasonable. This 'renormalization' revises Eq. (2) to read

$$Z = \left(\frac{(2i\pi)^{N-1}}{\prod_{j=1}^{N-1} a_j}\right)^{1/2} \prod_{j=1}^{N-1} \exp\left[-\frac{i}{2}\frac{\tilde{J}_j^2}{a_j}\right] \tag{15}.$$

Since $\vec{J}$ is defined via links we have characterized the symmetry amplitude in terms of relations and the non-zero eigenvalues of $\vec{\vec{A}}$.

To obtain $Z$ for our graphical example *supra*, we need the eigenvalues and their corresponding orthonormalized eigenvectors for $\vec{\vec{A}}$ of Eq. (10):

$$\langle 1| \equiv \langle a_1 = 5| = \begin{bmatrix} \frac{-1}{2\sqrt{3}} & \frac{1}{\sqrt{3}} & \frac{-1}{2\sqrt{3}} & \frac{1}{2\sqrt{3}} & \frac{-1}{\sqrt{3}} & \frac{1}{2\sqrt{3}} \end{bmatrix}$$

$$\langle 2| \equiv \langle a_2 = 3| = \begin{bmatrix} \frac{1}{2} & 0 & \frac{-1}{2} & \frac{-1}{2} & 0 & \frac{1}{2} \end{bmatrix}$$

$$\langle 3| \equiv \langle a_3 = 3| = \begin{bmatrix} \frac{-1}{2\sqrt{3}} & \frac{1}{\sqrt{3}} & \frac{-1}{2\sqrt{3}} & \frac{-1}{2\sqrt{3}} & \frac{1}{\sqrt{3}} & \frac{-1}{2\sqrt{3}} \end{bmatrix}$$

$$\langle 4| \equiv \langle a_4 = 2| = \begin{bmatrix} \frac{-1}{\sqrt{6}} & \frac{-1}{\sqrt{6}} & \frac{-1}{\sqrt{6}} & \frac{1}{\sqrt{6}} & \frac{1}{\sqrt{6}} & \frac{1}{\sqrt{6}} \end{bmatrix}$$

$$\langle 5| \equiv \langle a_5 = 1| = \begin{bmatrix} \frac{-1}{2} & 0 & \frac{1}{2} & \frac{-1}{2} & 0 & \frac{1}{2} \end{bmatrix}$$

$$\langle 6| \equiv \langle a_6 = 0| = \begin{bmatrix} \frac{1}{\sqrt{6}} & \frac{1}{\sqrt{6}} & \frac{1}{\sqrt{6}} & \frac{1}{\sqrt{6}} & \frac{1}{\sqrt{6}} & \frac{1}{\sqrt{6}} \end{bmatrix}$$

Since $\tilde{J}_i = \langle i|J\rangle$ we have

$$\frac{\tilde{J}_1^2}{a_1} = \frac{1}{5}\left(\frac{3e_1 + 4e_2 - 3e_3 + 2e_4 - 3e_5 + 3e_6 + 2e_7}{2\sqrt{3}}\right)^2$$

$$\frac{\tilde{J}_2^2}{a_2} = \frac{1}{3}\left(\frac{-e_1 - e_3 - 2e_4 + e_5 + e_6 + 2e_7}{2}\right)^2$$

$$\frac{\tilde{J}_3^2}{a_3} = \left(\frac{e_1 - e_3 + e_5 - e_6}{2}\right)^2$$

$$\frac{\tilde{J}_4^2}{a_4} = \left(\frac{-e_2 + e_4 + e_7}{\sqrt{3}}\right)^2$$

$$\frac{\tilde{J}_5^2}{a_5} = \left(\frac{e_1 + e_3 + e_5 + e_6}{2}\right)^2$$

to be summed and multiplied by –i/2 for the phase of *Z*. The amplitude is

$$\left(\frac{(2i\pi)^{N-1}}{\prod_{j=1}^{N-1} a_j}\right)^{1/2} = \left(\frac{32i\pi^5}{90}\right)^{1/2}$$

where N = 6 for our graph.

In summary: SCC ( $\vec{\vec{A}}\vec{v} \propto \vec{J}$ ) → actional ( $\Sigma = \frac{1}{2}\vec{\vec{A}} + \vec{J}$ ) → symmetry amplitude (*Z*) → relative probability for a particular spatio-temporal configuration over all possible configurations and outcomes of interest. This provides conceptually, if not analytically, a basis for the RBW ontology and methodology, enumerated as follows:

1. Each piece of equipment in an experimental set-up results from a large number of spatiotemporally dense relations, so low-intensity sources and high-sensitivity detectors must be used to probe the realm of rarefied relations described by QM (Figure 1).

2. A "detector click" is a subset of the detector that also results from a large number of spatiotemporally dense relations; we infer the existence of a rarified set of relations between the source and the detector at the beginning of the click's worldline.
3. It is this inferred, rarified set of relations for which we compute the symmetry amplitude.
4. A symmetry amplitude must be computed for each of all possible click locations (experimental outcomes) and this calculation must include (tacitly if not explicitly) all relevant information concerning the spacetime relationships (e.g., distances and angles) and property-defining relations (e.g., degree of reflectivity) for the experimental equipment.
5. The relative probability of any particular experimental outcome can then be determined by squaring the symmetry amplitude for each configuration (which includes the outcomes) and normalizing over all configurations.

*3.2 Two-source Propagator.* To obtain the QM transition amplitude between a single pair of 'sources' we need the spatially discrete and temporally continuous counterpart to Eq. (2). Therefore, we must find $D_{im}(t-t')$ in

$$Z(J) \propto \exp\left[\frac{-i}{2}\iint dt dt' J_i(t) D_{im}(t-t') J_m(t')\right] \tag{16}$$

where we've implied sums over repeated indices. $D_{im}(t-t')$ is given by

$$-\begin{pmatrix} m\dfrac{d^2}{dt^2}+k & k_{12} \\ k_{12} & m\dfrac{d^2}{dt^2}+k \end{pmatrix} D_{im}(t-t') = \begin{pmatrix} \delta(t-t') & 0 \\ 0 & \delta(t-t') \end{pmatrix} \tag{17}$$

Using

$$\delta(t-t') = \int \frac{d\omega}{2\pi} e^{i\omega(t-t')} \tag{18}$$

and assuming

$$D_{im}(t-t') = -\begin{pmatrix} \displaystyle\int \frac{d\omega}{2\pi} A(\omega) e^{i\omega(t-t')} & \displaystyle\int \frac{d\omega}{2\pi} B(\omega) e^{i\omega(t-t')} \\ \displaystyle\int \frac{d\omega}{2\pi} B(\omega) e^{i\omega(t-t')} & \displaystyle\int \frac{d\omega}{2\pi} A(\omega) e^{i\omega(t-t')} \end{pmatrix} \tag{19}$$

we find

$$A = \frac{\omega^2 m - k}{k_{12}^2 - \left(\omega^2 m - k\right)^2} \tag{20}$$

and

$$B = \frac{k_{12}}{k_{12}^2 - \left(\omega^2 m - k\right)^2} \tag{21}$$

so the QM transition amplitude in this simple case is given by

$$Z(J) \propto \exp\left[-\frac{i}{\hbar}\iint dt dt' J_1(t) D_{12} J_2(t')\right] = \exp\left[\frac{i}{\hbar}\iiint \frac{dt' dt d\omega}{2\pi}\frac{J_1(t) k_{12} e^{i\omega(t-t')} J_2(t')}{k_{12}^2 - (\omega^2 m - k)^2}\right] \tag{22}$$

having restored $\hbar$, used $D_{12} = D_{21}$ and ignored the "self-interaction" terms $J_1 D_{11} J_1$ and $J_2 D_{22} J_2$. We can simplify the expression via the Fourier transform

$$j_2(\omega) \equiv \int J_2(t') e^{-i\omega t'} dt' \tag{23}$$

so that

$$Z(J) \propto \exp\left[\frac{i}{\hbar}\iint \frac{dt d\omega}{2\pi}\frac{J_1(t) k_{12} e^{i\omega t} j_2(\omega)}{k_{12}^2 - (\omega^2 m - k)^2}\right] \tag{24}$$

With

$$J_1(t) = \int j_1(\omega') e^{i\omega' t}\frac{d\omega'}{2\pi} \tag{25}$$

we have

$$Z(j) \propto \exp\left[\frac{i}{\hbar}\iiint \frac{d\omega' dt d\omega}{(2\pi)^2}\frac{j_1(\omega') k_{12} e^{it(\omega+\omega')} j_2(\omega)}{k_{12}^2 - (\omega^2 m - k)^2}\right] \qquad (26).$$

Using Eq. (18) we find

$$Z(j) \propto \exp\left[\frac{i}{\hbar}\int \frac{d\omega}{2\pi}\frac{j_1(-\omega) k_{12} j_2(\omega)}{\left(k_{12}^2 - (\omega^2 m - k)^2\right)}\right] \tag{27}$$

or

$$Z(j) \propto \exp\left[\frac{i}{\hbar}\int \frac{d\omega}{2\pi}\frac{j_1(\omega)^* k_{12} j_2(\omega)}{\left(k_{12}^2 - (\omega^2 m - k)^2\right)}\right] \tag{28}$$

with $J_1$(t) real.

*3.3 Twin-Slit Amplitude.* We now use the amplitude of subsection 3.2 to analyze the twin-slit experiment. There are four *J*'s which must be taken into account when computing the amplitude (Figure 4), so we will use the solution obtained in subsection 3.2 to link $J_1$ with each of $J_2$ and $J_4$, and $J_3$ with each of $J_2$ and $J_4$, i.e., $J_1 \leftrightarrow J_2 \leftrightarrow J_3$ and $J_1 \leftrightarrow J_4 \leftrightarrow J_3$. In doing so, we ignore the contributions from other pairings, i.e., the exact solution would contain one integrand with $J_n \rightarrow J_i(t)$, i = 1,2,3,4. Also, we're finding interference effects while ignoring diffraction effects, i.e., a precise solution would employ two *J*'s for each slit–one *J* for each edge of each slit. [This is not an issue if one uses laser-excited atoms in lieu of "slits" (Scully & Druhl, 1982).]

**Figure 4**

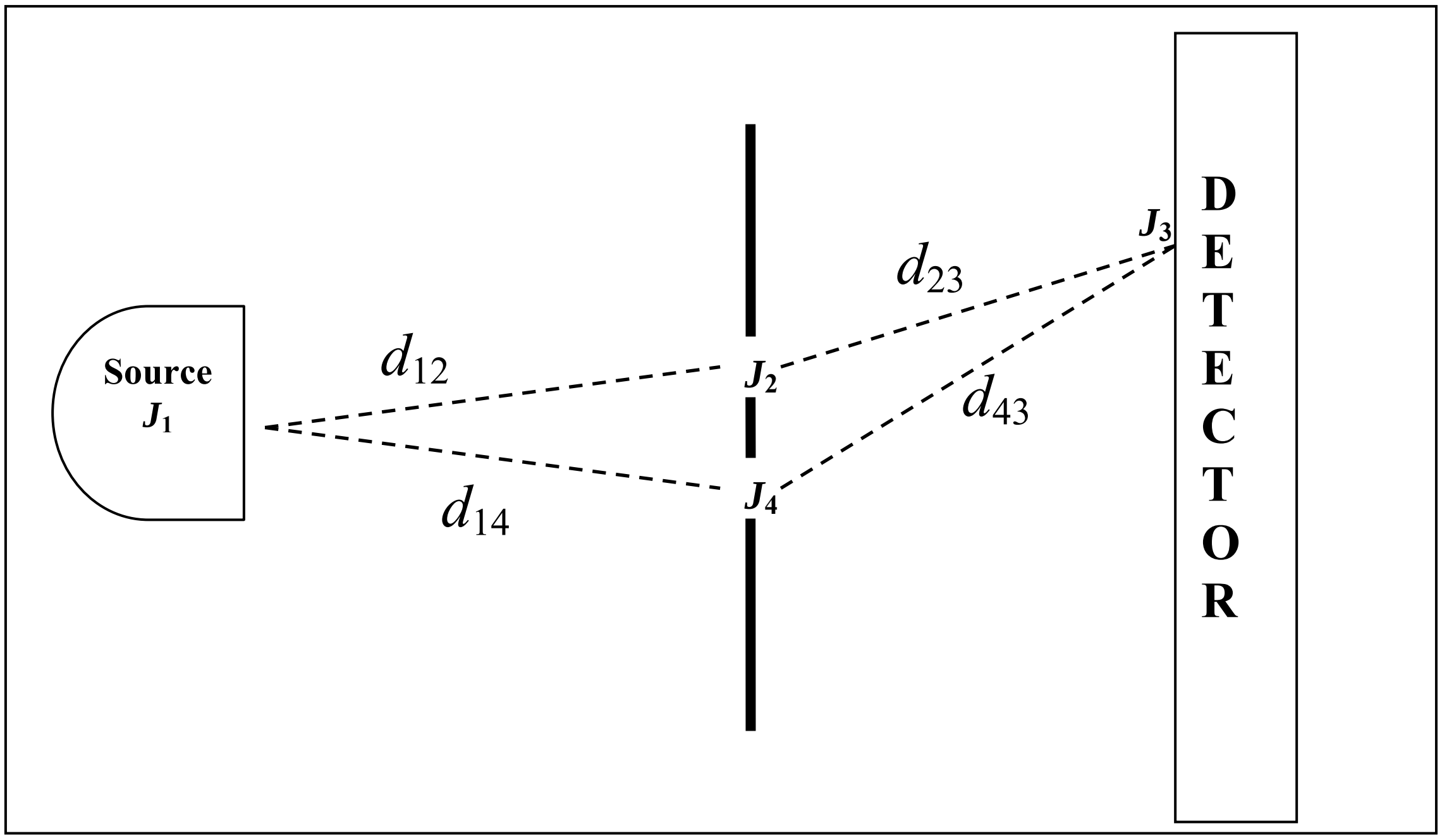

Finally, we assume a monochromatic source of the form $j_1(\omega)^* = \Gamma_1\delta(\omega-\omega_o)$ with $\Gamma_1$ a constant, so the amplitude between $J_1$ and $J_2$ is

$$Z(j) \propto \exp\left[\frac{i}{2\pi\hbar}\frac{\Gamma_1 k_{12} j_2(\omega_o)}{\left(k_{12}^2-(\omega_o^2 m-k)^2\right)}\right] \tag{29}$$

whence we have for the amplitude between $J_1$ and $J_3$ via $J_2$ and $J_4$

$$\psi \propto \exp\left[\frac{i}{2\pi\hbar}(\Gamma_1 d_{12} j_2+\Gamma_2 d_{23} j_3)\right]+\exp\left[\frac{i}{2\pi\hbar}(\Gamma_1 d_{14} j_4+\Gamma_4 d_{43} j_3)\right] \tag{30}$$

where

$$d_{im} = \frac{k_{im}}{\left(k_{im}^2 - (\omega_o^2 m - k)^2\right)} \tag{31}$$

with $\psi$ the QM symmetry amplitude. [$Z$ corresponds to the QM propagator which yields the functional form of $\psi$ between spatially localized sources, as will be seen below.] With the source equidistance from either slit (or laser-excited atoms instead of "slits") we expect the phase $\Gamma_1\, d_{12}j_2$ equals the phase $\Gamma_1\, d_{14}j_4$, i.e., $J_2$ and $J_4$ are coherent, so we have the familiar form

$$\psi \propto \exp\left[\frac{i}{2\pi\hbar}(\Gamma_2 d_{23} j_3)\right] + \exp\left[\frac{i}{2\pi\hbar}(\Gamma_4 d_{43} j_3)\right] \tag{32}$$

*3.4 Non-Separable Spatial Distance in Schrödinger Wave Mechanics.* In order to illustrate non-separable spatial distance in the Schrödinger formalism, we simply compare the Schrödinger twin-slit wavefunction to Eq. (32). The free-particle propagator in wave mechanics is (Shankar, 1994)

$$U(x_2, t; x_1, 0) = \sqrt{\frac{m}{2\pi\hbar i t}} \exp\left[\frac{im(x_2 - x_1)^2}{2\hbar t}\right] \tag{33}$$

for a particle of mass $m$ moving from $x_1$ to $x_2$ in time $t$. This 'exchange' particle has no dynamic counterpart in the formalism used to obtain Eq. (32), but rather is associated with the oscillatory nature of the spatially discrete 'source' (see below). According to our view, this propagator is tacitly imbued "by hand" with notions of dynamical/diachronic entities, space and time per its derivation via the free-particle Lagrangian. In short, the construct of this propagator bypasses explicit, self-consistent construct of trans-temporal objects, space and time thereby ignoring the self-consistency criterion fundamental to the action. The self-inconsistent, tacit assumption of a single particle with two worldlines (a "free-particle propagator" for each slit) is precisely what leads to the "mystery" of the twin-slit experiment. This is avoided in our formalism because $Z$ does not represent the propagation of a particle between 'sources', e.g., $q_i(t) \neq x(t)$. Formally, the inconsistent, tacit assumption is reflected in $-\frac{1}{2}\vec{Q}\cdot\vec{\vec{A}}\cdot\vec{Q} \rightarrow \int\left(\frac{m}{2}\dot{x}^2\right)dt$ where ontologically $m$ (which is *not* the same $m$ that appears in our oscillator potential) is the mass of the 'exchange' particle (i.e., purported dynamical/diachronic entity moving between 'sources' – again, the ontic status of this entity is responsible for the "mystery") and $x$(t) (which, again, is *not* equal to $q_i$(t)) is obtained by *assuming* a particular spatial metric (this assumption *per se* is not responsible for the "mystery"). Its success in producing an acceptable amplitude when integrating over all paths $x$(t) in space ('wrong' techniques can produce 'right' answers), serves to deepen the "mystery" because the formalism, which requires interference between different spatial paths, is not consistent with its antecedent ontological assumption, i.e., a single particle taking two paths causes a single click or a 'matter wave' distributed throughout space causes a spatially localized detection event. There is no such self-inconsistency in our approach, because $Z$ is not a "particle propagator" but a 'mathematical machine' which measures the degree of "symmetry" contained in the "all at once" configuration of dynamical/diachronic entities, space and time represented by the actional, as explained *supra*. Thus, this QM "mystery" results from an attempt to tell a dynamical story in an adynamical situation. Continuing, we have

$$\psi(x_2,t) = \int U(x_2,t;x',0)\psi(x',0)dx' \tag{34}$$

and we want the amplitude between sources located at $x_1$ and $x_2$, so $\psi(x',0) = \alpha\delta(x'-x_1)$ whence

$$\psi_{12} = \alpha\sqrt{\frac{m}{2\pi\hbar it}}\exp\left[\frac{imx_{12}^2}{2\hbar t}\right] = \alpha\sqrt{\frac{m}{2\pi\hbar it}}\exp\left[\frac{ipx_{12}}{2\hbar}\right] \tag{35}$$

where $x_{12}$ is the spatial distance between sources $J_1$ and $J_2$, $t$ is the interaction time and $p = \frac{mx_{12}}{t}$. Assuming the interaction time is large compared to the 'exchange' particle's characteristic time so that $x_{12}$ is large compared to $\frac{\hbar}{p}$ we have

$$\psi = \psi_{23} + \psi_{43} \propto \exp\left[\frac{ipx_{23}}{2\hbar}\right] + \exp\left[\frac{ipx_{43}}{2\hbar}\right] \tag{36}$$

for the Schrödinger twin-slit wavefunction. Comparing Eqs. (32) and (36), we infer

$$\frac{p}{2\hbar}x_{ik} = \frac{\Gamma_i d_{ik} j_k}{2\pi\hbar} \tag{37}$$

Assuming the impulse $j_k$ is proportional to the momentum transfer $p$, we have

$$x_{im} \propto \frac{\Gamma_i k_{im}}{\left(k_{im}^2 - (\omega_o^2 m - k)^2\right)} \tag{38}$$

relating the spatial separation $x_{im}$ of the trans-temporal objects $J_i$ and $J_m$ to their intrinsic ($m$, $k$, $\omega_o$) and relational ($k_{im}$) ‘dynamical’ characteristics.

While Eq. (38) suggests a relationship between the spacetime metric and dynamics *a la* general relativity, $x_{im}$ is distinct from $\overset{\Rightarrow}{g}(\vec{e}_i, \vec{e}_m)$, where $\{\vec{e}_j\}$spans the tangent space T of the spacetime manifold and $\overset{\Rightarrow}{g} \in T^* \otimes T^*$ is the spacetime metric with T* dual to T. The spatial separation of Eq. (38) exists only between interacting ($k_{im} \neq 0$) dynamical/diachronic entities, in stark contrast to the field $\overset{\Rightarrow}{g}(\vec{e}_i, \vec{e}_m)$ which takes on values for all points of the differentiable spacetime manifold, even in regions where the stress-energy tensor is zero. According to RBW, there is no mediating particle or wave (of momentum $p$ or otherwise) moving ‘through space’ from the source to the detector.

Again, per RBW, reality is fundamentally described relationally throughout space and time. A subset of all physically realizable RBW descriptions lends itself to dynamical interpretation/storytelling, but not all. The “mystery” of the twin-slit experiment is deflated by realizing that it is simply a phenomenon whose description resides outside that dynamical subset. And, as Feynman suspected, this is the case with all “mysterious” QM phenomena – they are totally explicable via RWOT, but dynamical reality is a proper subset of a spatiotemporally holistic reality and “mysterious” QM phenomena are simply not elements of that dynamical subset. We finish with an introduction to the quantum gravity regime of the RBW unification scheme.

**4. THE QUANTUM GRAVITY REGIME**

Neither Newtonian gravity nor general relativity suffices to account exhaustively for intra/inter-galactic motion given the matter/energy observed. As Smolin (2006) points out, these motions seem to require additional attractive/repulsive forms of matter/energy, respectively. However, there are no direct observations of this matter/energy, so it’s called “dark.” What does our proposed unification scheme say about the nature of dark matter and dark energy? We do not pretend to have an answer to this question, but we do offer a guess as to where it will be found. The phenomena associated with dark matter/energy are actually part of a new regime in our unification scheme, i.e., one set of continuous time, continuous space, dense relations joined to another set of continuous time, continuous space, dense relations by rarefied relations. Since observational data concerning galactic motion is obtained by spectral analysis, which falls in the realm of continuous time, discrete space and rarefied relations, while the dynamical/diachronic entities in question are described in the realm of continuous time, continuous space and dense relations, we are dealing with an uncharted regime of physics per Figure 1. It remains a technical challenge in the context of our proposed formalism to describe this new regime, which we would call 'true' QG.

Interestingly, this 'true' QG regime is not characterized by large mass-energy densities (dense relations) as commonly assumed, but by a mix of dense *and* rarefied relations. In the context of a fundamentally discrete theory of spacetime and stress-energy, so-called "physical" singularities associated with the centers of black holes and the big bang are understood to result from approximations made in the limits of temporal and spatial continuity, viz., a large but denumerable number of elements are rendered indenumberable to obtain worldlines in the limit of temporal continuity and small but non-zero spatial separations between worldlines render zero spatial volume in the limit of spatial continuity. Since these approximations are introduced for computational purposes, the resulting singularities are best described as *mathematical* not physical. Indeed, one sees immediately that the limit of spatial volume → zero does not produce infinities in the graphical approach; instead, one simply loses the multiplicity (of worldlines) afforded by spatiality. Thus, the current 'faux' QG regime poses no conceptual challenge for our formalism; indeed it does not even constitute a new regime.

## 5. CONCLUSION

We have presented heuristically a formalism for the unification of physics which is motivated by, and inextricably linked to, our interpretation of QM, just as Smolin (2006) suspected would be the case. The "utterly simple idea" at the bottom of this unification scheme is that dynamical/diachronic entities (obtained via trans-temporal identification), space and time cannot be defined independently of one another, which is also consistent with Smolin's prediction that "the key" to unification would be "the nature of time." To codify this demand for self-consistency, we proposed a self-consistency criterion (SCC) in the context of discrete graph theory *a la* Wise (2006) that underlies the discrete action. To do this, we first constructed the 'relationship' matrix for the source-free field portion of the discrete action from a boundary operator on the spacetime chain complex of the graph so that it was proportional to its counterpart in the action for coupled harmonic oscillators on discrete spacetime. Defining the discrete source vector relationally via links of the graph then resulted in an SCC fundamental to the discrete action. As predicted by Toffoli (2003), our basis for the action (the SCC) results from a mathematical tautology, viz., the topological maxim "the boundary of a boundary is zero," which guarantees the consistency of fields and divergence-free sources in general relativity and electromagnetism. This result illustrates how 'multiplicity' in dynamism might arise from the fact that the SCC constrains sets of dynamical constants *as a whole, rather than in isolation*. Further, the SCC requires a choice of spacetime metric over the graph, which means we've a 'dynamic' spacetime structure *a la* GR. This approach constitutes a *unification* of physics as opposed to a mere *discrete approximation* thereto, since we are proposing a basis for the action, which is otherwise fundamental. Thus, we believe this approach to unification is worthy of further scrutiny.